# Pressure-induced isostructural phase transition and correlation of FeAs coordination with the superconducting properties of 111-type Na1-xFeAs

Qingqing Liu[1], Xiaohui Yu[1,2], Xiancheng Wang[1], Zheng Deng[1], Yuxi Lv[1], Jinlong Zhu[1], Sijia Zhang[1], Haozhe Liu[3], Wenge Yang[4], Lin Wang,[4] Hokwang Mao[4,5], Guoyin Shen[5], Zhongyi Lu[6], Yang Ren[7], Zhiqiang Chen[8], Zhijun Lin[2], Yusheng Zhao[2], Changqing Jin[1]

[1]Institute of Physics, Chinese Academy of Sciences, Beijing 100190, China

[2]LANSCE, Los Alamos National Laboratory, Los Alamos, NM 87545, USA

[3]Natural Science Research Center, Harbin Institute of Technology, Harbin 150080, China

[4]HPSynC, Geophysical Laboratory, Carnegie Institution of Washington, Argonne, IL 60439

[5]HPCAT, Geophysical Laboratory, Carnegie Institution of Washington, Argonne, IL 60439

[6]Department of Physics, Renmin University, Beijing 100872, China

[7]Advanced Photon Source, Argonne National Laboratory, Argonne, IL 60439

[8]National synchrotron light source, Brookhaven National Lab, Upton, NY 11973

Email address of corresponding author: Jin@iphy.ac.cn

## Abstract

The effect of pressure on the crystalline structure and superconducting transition temperature ($T_c$) of the 111-type Na1-xFeAs system using in situ high pressure synchrotron x-ray powder diffraction and diamond anvil cell techniques is studied. A pressure-induced tetragonal to tetragonal isostructural phase transition was found. The systematic evolution of the FeAs4 tetrahedron as a function of pressure based on Rietveld refinements on the powder x-ray diffraction patterns was obtained. The non-monotonic $T_c(P)$ behavior of Na1-xFeAs is found to correlate with the anomalies of the distance between the anion (As) and the iron layer as well as the bond angle between As-Fe-As for the two tetragonal phases. This behavior provides the key structural information in understanding the origin of the pressure dependence of Tc for 111-type iron pnictide superconductors. A pressure-induced structural phase transition is also observed at 20 GPa.

# Introduction

The recent discovery of superconductivity at 26 K of $LaO_{1-x}F_xFeAs$ [1] opened a new door for research in the area of high-temperature superconductors. Superconductivity with different critical temperatures has been discovered in several families of layered iron pnictides including 1111, 122, 111-iron arsenides, or 11-selenides [2–14]. Pressure is important in the study of iron pnictide superconductors because it directly tunes the electronic configuration as demonstrated by several typical iron pnictide compounds [15–25]. In Fe-based superconductors, the effect of pressure on the superconducting transition temperature (Tc) is complex and depends sensitively on the composition of the materials. Therefore, the correlation between the pressure-tuned superconductivity and the atomic structure under pressure plays a key role in the search for new materials as well as in the elucidation of the mechanism of superconductivity in iron arsenide superconductors. The "111" system is the simplest for the iron arsenic superconductors. Therefore the knowledge about this system would be more straightforward to understand the superconducting mechanism or help designing new iron based superconductors. However there are very rare reports on "111" system comparing to "1111" or "122" since the samples contain very hygroscopic alkaline metal. Recently, we have reported on the effect of pressure on the superconductivity of 111-type Na1-xFeAs that crystallizes into the same structure[26] as that of LixFeAs superconductor[8]. We found that the superconducting critical temperature of Na1-xFeAs can reach a maximum of 31 K at approximately 3 GPa before the Tc decreases at a higher pressure. This pressure effect on the 111-type Na1-xFeAs is similar to that on the 1111-type LaFeAsO [16] and the 122-type $AFe_2As_2$ (A = Sr, Ba) systems [17, 18]. However, the effect is totally different from that for the isostructural 111-type LixFeAs where Tc is suppressed linearly with pressure [21–23]. To provide insights into the pressure behavior of the 111-type $Na_{1-x}FeAs$, studies on crystal structural evolution as a function of pressure based on in situ high-pressure synchrotron x-ray powder diffraction data and Rietveld refinement are performed in this work. A pressure-induced tetragonal to tetragonal isostructural phase transition was observed at approximately 3 GPa. The non-monotonic $T_c(P)$ behavior of $Na_{1-x}FeAs$ was found to correlate with the drastic changes in FeAs coordination such as the anion (As) height from the iron layer and the As-Fe-As bond angle, which indicates the key role of FeAs tetrahedron geometry in modulating electronic properties. A pressure-induced phase transition at 20 GPa was also observed.

## Experimental Section

The $Na_{1-x}FeAs$ polycrystalline sample was prepared by the solid-state reaction method using high-purity $Na_3As$, Fe, and As powders as starting materials, similar to the fabrication of $Li_xFeAs$ reported elsewhere [8, 9]. The in situ high-pressure angle dispersive x-ray diffraction (XRD) method is utilized for the structural studies of the $Na_{1-x}FeAs$ powder samples. The experiments were performed at several synchrotron facilities, including the High Pressure Collaborative Access Team of Advanced Photon Source (HPCAT) at the Argonne National Laboratory with $\lambda=0.368$ Å and the National Light Synchrotron Source (NSLS) with $\lambda= 0.413$ Å at the Brookhaven National Laboratory using diamond anvil cells with 500 μm cullet diamonds. Silicone oil, which was used as the pressure medium, provided the fine quasi-hydrostatic pressure environment within the pressure scope in the present study. Given that the $Na_{1-x}FeAs$ sample is very sensitive to air and moisture, the sample loading was conducted in a glove box filled with high purity Argon. The details of the specimen preparation can be found in Ref [27]. The diffraction patterns were collected using a Mar345 image-plate detector. The intensity versus 2 theta patterns was obtained using

FIT2D software. Data analysis of the diffraction profiles was performed using the GSAS-EXPGUI package [28]. Electrical resistivity measurements under high pressures were performed by a standard four-probe technique in the diamond anvil cell. MgO fine powder was used either as the insulating layer of the T301 stainless steel gasket or as the pressure-transmitting medium for the electric measurements. For all the above experiments, the gaskets were pre-indented from the original 300 to 100 μm in thickness. In these experiments, the pressure was applied at room temperature and measured by the ruby fluorescence method [29]. Our previous experiments indicated that pressure change little from room temperature to liquid helium temperature for the beryllium copper diamond anvil cell that is what we used in the present experiments.

## Results and Discussion

Figure 1 shows the XRD patterns of $Na_{1-x}FeAs$ at high pressure values (a) and its crystalline structure with the FeAs plane highlighted (b). A structural phase transition is clearly observed at 20 GPa as shown by the new peaks. To investigate the effect of pressure on the structural properties of the sample at atomic level, the XRD patterns of $Na_{1-x}FeAs$ below 20 GPa were analyzed with Rietveld refinements using the GSAS program package.[28] A typical refinement result at 11.1 GPa is shown in Figure 1(a) in which the fitted residuals $R_p$ and $R_{wp}$ were 5.16% and 6.38%, respectively. Figure 2 illustrates the pressure dependence of the unit cell parameters below 20 GPa based on the Rietveld refinements. The basal lattice parameter $a$ in $Na_{1-x}FeAs$ contracts by 3.6% below 20 GPa, whereas the lattice parameter $c$ drops by 4.7%, exhibiting anisotropic compression. Figure 2(b) exhibits the changes in volumes below 2.1 GPa and above 5.6 GPa as indicated by the plateau, signifying the isostructural phase transition from a tetragonal to a collapsed tetragonal phase since either no new peak appeared or the old peak disappeared. The pressure-induced isostructural phase transitions to a collapsed tetragonal were previously observed in 1111-type NdFeAsO [27] and 122-type CaFe2As2 [30]. The isostructural phase transition is related to the shearing movement of the charge reservoir layer during compression [27]. The present work provides a new example for the 111 system. As indicated by the results from the Rietveld refinements hereafter the isostructural phase transition is related to the drastic changes in FeAs4 coordination geometry. Figure 2(b) presents the unit-cell volume as a function of pressure in which the solid lines are the fitting results using the second-order Birch equation of state (EoS). With $B_0'$ fixed as 4, the ambient pressure isothermal bulk modulus $B_0$ =52.3(2) GPa below 2.1 GPa and $B_0$ =62.7(5) GPa above 5.6 GPa were obtained. These values are comparable to those of LixFeAs (57.3(6) GPa) [23] but significantly smaller than those of LaFeAsO0.9F0.1 (78.2 GPa)[31] and NdFeAsO0.88F0.12 (102.2 GPa) [27]. The results indicate that the charge reservoir of the alkaline metal for the 111-system is much softer than those of rare earth oxides for the 1111-system. This condition resulted in isostructural phase transition at a low pressure. Table 1 lists the compressibility and bulk modulus for several Fe-based superconductors for comparison. At about 16.8 GPa, the basal lattice parameter $a$ in Na1-xFeAs contracts by 3.6%, which is comparable to that of LixFeAs (3.9% at 10 GPa) but much larger than those of LaFeAsO0.9F0.1 (2.2% at 10 GPa) [31] and NdFeAsO0.88F0.12 (1.3% at 10 GPa) [27].

From the structural point of view, the FeAs4 tetrahedron in Na1-xFeAs is almost regular with an $\alpha$ and $\beta$ of 108.2° (×2) and 110.1° (×4), respectively, at ambient pressure. The angles are also closer to the values of 109.4° observed in layered iron arsenides at the peak values of Tc [32]. In contrast, the $\alpha$ and $\beta$ are 102.9° (×2) and 112.9° (×4) in the isostructural LixFeAs superconductor, respectively, which gives the most compressed FeAs4 tetrahedron in the basal plane among the iron arsenide superconductors. The application of pressure can significantly affect the crystalline structure and electronic properties of Na1-xFeAs through tuning the geometric structure of FeAs coordination. The anion heights from the iron layer and the As-Fe-As bond angles (a 2-fold $\alpha$ angle bisected by the $c$ axis) of the FeAs4 tetrahedra in Na1-xFeAs at various pressure values are calculated using Rietveld refinements. Figure 3 illustrates the pressure dependence of the anion height and As-Fe-As bond angles, respectively. The As-Fe-As bond angle evidently increases with an increase in pressure, resulting in a peak value, and then rapidly decreases at a high pressure. The pressure dependence of the anion height is contrary to that of the As-Fe-As bond angle. Structural analysis shows that this behavior is attributed to the change in compression of the FeAs4 tetrahedra at different applied pressure values. At a low pressure, FeAs4 tetrahedra present greater compression along the $c$-axis direction than in the basal plane and tend to approach an almost regular FeAs4 tetrahedra of 109.4°. Therefore, the compression of Na1-xFeAs is expected to be accommodated by the change in the softest parts of the structure, i.e., at the charge reservoir block of the double Na layers. The change in distance $h$ between the Fe layer and arsenic reflects this tendency. The fast decrease in the $c$ axis over the $a$ axis increases the As-Fe-As angle but at the same time reduces $h$, which is observed in the low pressure range of Fig. 3. This evolution results in isostructural phase transition from one tetragonal to a collapsed tetragonal phase, which is attributed to the shear movements of the charge reservoir layer. With a further increase in pressure, compression in the basal plane of the FeAs4 tetrahedra becomes significantly greater than that along the $c$-axis direction, resulting in a rapid decrease in the As-Fe-As bond angles (2-fold $\alpha$ angle bisected by the $c$ axis) of the FeAs4 tetrahedra. This in turn distorts the FeAs4 tetrahedra away from the regular shape. Remarkably, the pressure range at which the maximum T$c$ as shown in Figure 3(b) is found coincides with the onset of the FeAs4 tetrahedron geometry change and the discontinuous change of the crystal cell volume. Within the transitional region, the FeAs4 tetrahedron shape is closer to the regular shape with an average As-Fe-As angle of 109.4°. The anion height is 1.38 Å, which is far less than 1.42 Å at ambient pressure. With an increase in pressure, the As-Fe-As bond angle decreases and dramatically deviates from the ideal tetrahedron value of 109.47°. The anion height also increases with an increase in pressure. This behavior strongly suggests that the pressure-induced higher Tc superconducting phenomenon is associated with the evolution of the distorted FeAs4 tetrahedron as it approaches the regular tetrahedron and optimized anion height. With an increase in pressure, the negative pressure coefficients of Tc are observed, which are rationalized in terms of the increased tetrahedral distortion away from the regular shape at a higher pressure. This is somehow different from those observed in Ca122 where the parent compound becomes superconductive in certain applied pressure region but with a sharp boundary to the non superconducting high pressure phase [33]. Assuming that pressure increases Tc of the first tetragonal phase while it suppresses Tc of the second tetragonal phase for the Na 111, one can however further eventfully correlate Tc with the change of FeAs4 geometry as revealed in our studies. Therefore the FeAs4 geometry is the key factor that determines the superconducting transition temperature.

The superconductivity is determined physically by the Fermi surface and the interaction that is responsible for mediating the electrons from the Fermi surface to form Cooper pairs. The electron-phonon interaction is unlikely to carry such mediation in iron pnictides [34] because the calculated Tc is too low. On the other hand, electron pairing takes place between the hole-type Fermi sheet around the Gamma point and the electron-type Fermi sheet around the M point mediated by involving the antiferromagnetic superexchange interaction J2 between the next nearest neighbors

Fe-Fe atoms[35~38] as shown in Figure 1(b). The J2 superexchange interaction is bridged by the As atom through the covalent bonding between the Fe and As atoms[36]. Therefore, this interaction is sensitive to the local geometry of Fe-As bonding. Here pnictide superconductors are multi-orbital systems. For an FeAs tetrahedral layer, the Fe atoms form a square lattice as shown in Fig.1b, with two As atoms above while the other two below. The metallic property of pnictides is determined by the Fe-Fe square lattice, specifically by the direct overlap between $d_{x^2-y^2}$ orbitals of the nearest neighbor Fe atoms, while the Fe-As bonding happens between $d_{xz}$ ($d_{yz}$) orbital of iron and $p_x$($p_y$) orbital of arsenic. Thus as the FeAs4 tetrahedron goes regular namely the As-Fe-As angle alpha approaches 109.47 degree, the superexchange interaction J2 as shown in Fig. 1b goes largest, meanwhile the Fermi surface and the density of states at the Fermi energy remain almost unchanged. The results of synchrotron x-ray diffraction indicate that the crystalline structure remains stable at least below 20 GPa. Therefore, the superconductivity evolution observed in the present work is merely caused by the changes in electronic structure at a high pressure, as revealed by the discontinuous change in crystal cell volume. According to the experimental results in Ref. [32], the ideal As-Fe-As bond angle is $\alpha = \beta = 109.47°$, which corresponds to the highest Tc of the 1111-type LnFeAsO system. A recent study showed that the Tcs for most Fe-based superconductors closely depend on the anion height (h) and exhibit a parabolic-type correlation between Tc and *h* with a maximum Tc around *h*=1.38 Å [39]. Interestingly, the superconductivity with the maximum Tc in Na1-xFeAs lies in the tetrahedron transitional region where both anion height (1.38 Å) and FeAs4 coordination are near the optimal values. This result further supports the concept that the regular FeAs4 tetrahedron and the optimal anion height, which are often coupled as shown in the present work, are the necessary crystal geometries for obtaining the maximum Tc for a given iron arsenide superconductor.

## Conclusions

The structure and superconductivity of 111-type Na1-xFeAs were investigated under high pressure using synchrotron x-ray powder diffraction and diamond anvil cell technique. The measurements of synchrotron x-ray diffraction show that Na1-xFeAs exhibits the ambient pressure isothermal bulk modulus $B_0$ =52.3(2) GPa below 2.1 GPa and $B_0$ =62.7(5) GPa above 5.6 GPa, which is comparable to that in isostructural LixFeAs. There is a pressure-induced isostructural phase transition at ~3 GPa as revealed by the discontinuous change in crystal cell volume and the evolution of FeAs4 tetrahedron shape with pressure. The structural evolution with applied pressure reveals an intimate link between the As-Fe-As bond angle and the anion height as well as the pressure-tuned superconductivity of the Na1-xFeAs superconductor. The results of the present study indicate the non-monotonic relation between $T_c(P)$ and the change in anion height and As-Fe-As bond angle with the maximum Tc correlating both with the regular FeAs4 tetrahedron and the optimal distance between the arsenic element and the iron layer.

**Acknowledgment**. This work was supported by NSF & MOST through research projects. HPSynC is supported as part of EFree, an Energy Frontier Research Center funded by the U.S. Department of Energy under Award DE-SC0001057. HPCAT is supported by DOE-BES, DOE-NNSA, and National Science Foundation

**Supporting Information Available**: Complete Ref. 30. This material is available free of charge via the Internet at http://pubs.acs.org/."

# Figure captions:

Figure 1. (a) XRD patterns of Na1-xFeAs at high pressure values. A structural phase transition appears at 20 GPa (the circle indicates the peaks from the new structure). The XRD pattern at 11.1 GPa is refined using the Rietveld method; (b) The schematic view of the 111-type crystalline structure of NaFeAs with the FeAs tetrahedron geometry is highlighted in the FeAs plane.

Figure 2. Pressure dependence of the lattice parameters (a) and unit cell volume (b) of Na1-xFeAs. The solid lines in (b) are the fitting results according to second-order Birch EoS.

Figure 3. (a) Pressure dependence of As-Fe-As bond angles ($\alpha$ angle) and anion height from the iron layer; (b). Pressure dependence of Tc for Na1−xFeAs obtained from resistance measurements. Points are experimental data, whereas the lines are polynominal fit to the experiment; (c). Phase diagram of the crystalline structure change of coordination geometry and superconducting Tc as a function of pressure.

**Table 1**. The compressibility and bulk modulus for several iron-based superconductors for comparison.

# Table 1

| Compound | $r_a$ (*a*- axis compressibility) (%) | $r_c$ (*c*-axis compressibility) (%) | $r_c/r_a$ | Bulk modulus (GPa) |
|---|---|---|---|---|
| Na1-xFeAs [this work] | 3.6 | 4.7 | 1.3 | 52.3(2) 62.7(5) |
| LixFeAs [21] | 3.9 | 5.5 | 1.4 | 57.3(6) |
| LaFeAsO0.9F0.1 [31] | 2.2 | 4.4 | 2.0 | 78(2) |
| NdFeAsO0.88F0.12 [27] | 1.3 | 5.5 | 4.2 | 102(2) |

**Fig.1(a)**

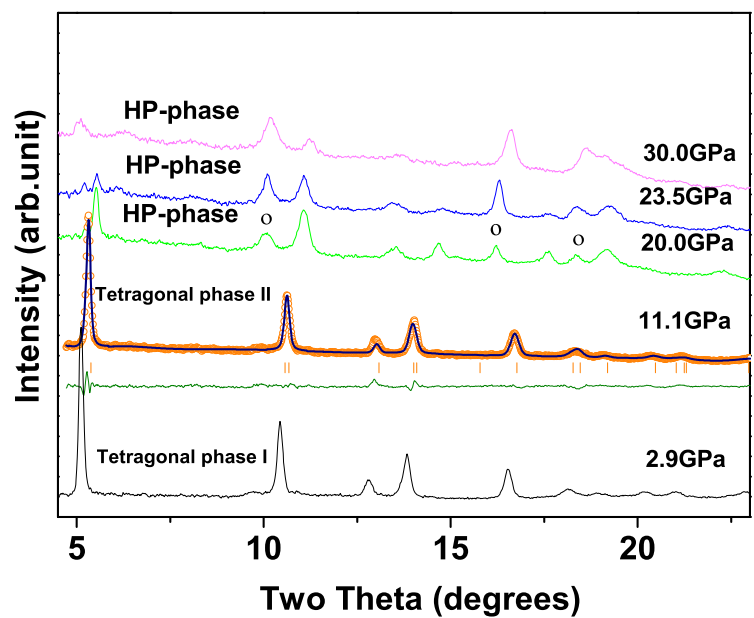

**Fig.1(b)**

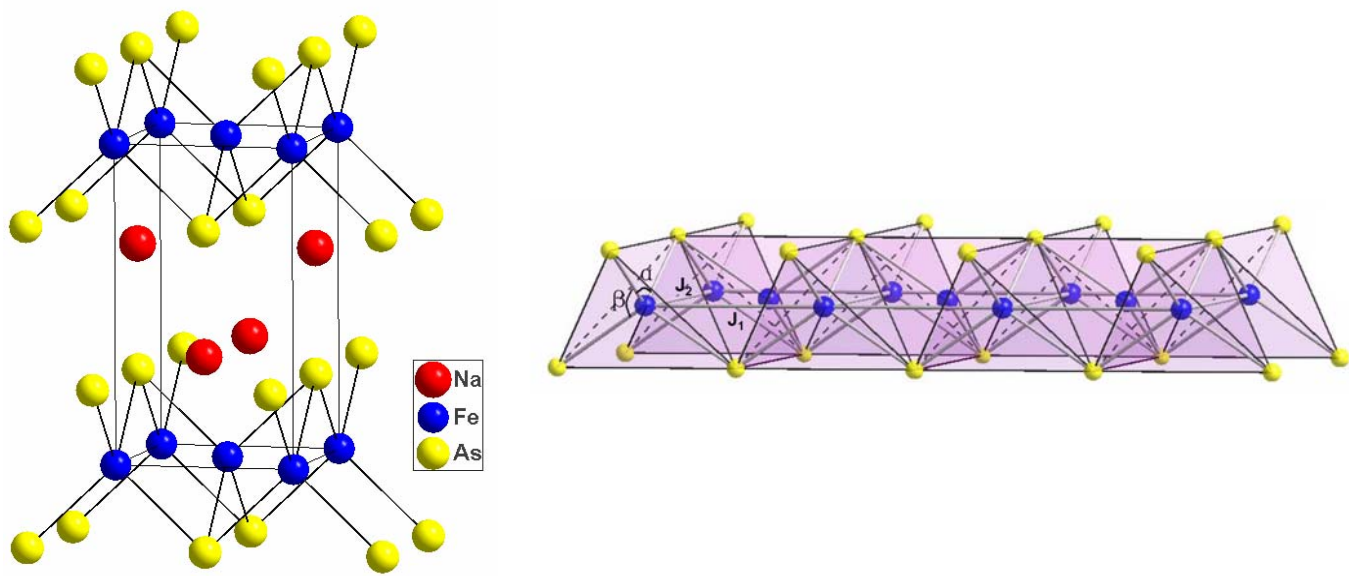

**Fig. 2(a)**

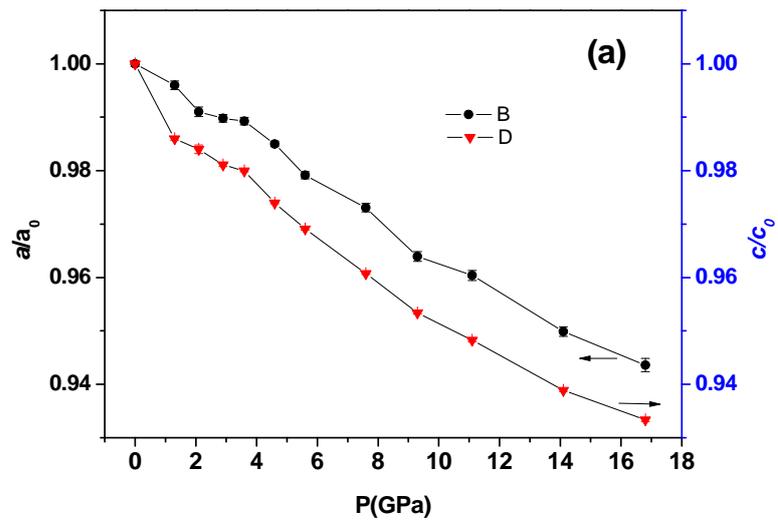

**Fig. 2(b)**

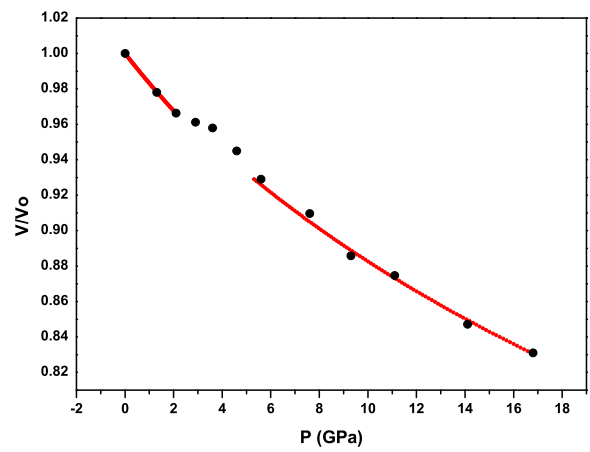

**Fig.3 (a)**

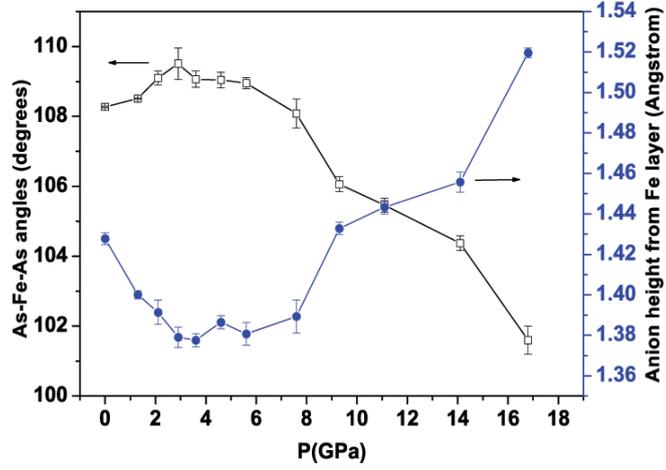

**Fig.3 (b)**

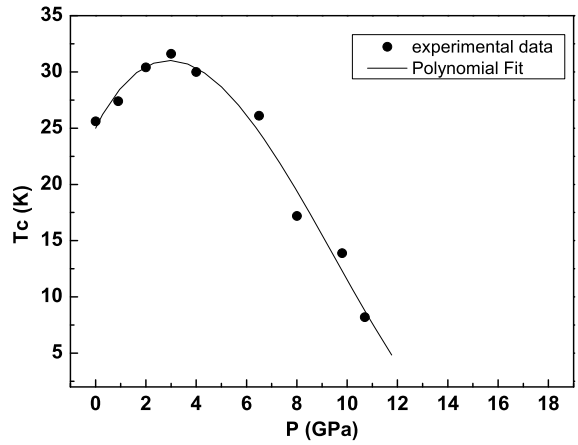

**Fig.3 ( c)**

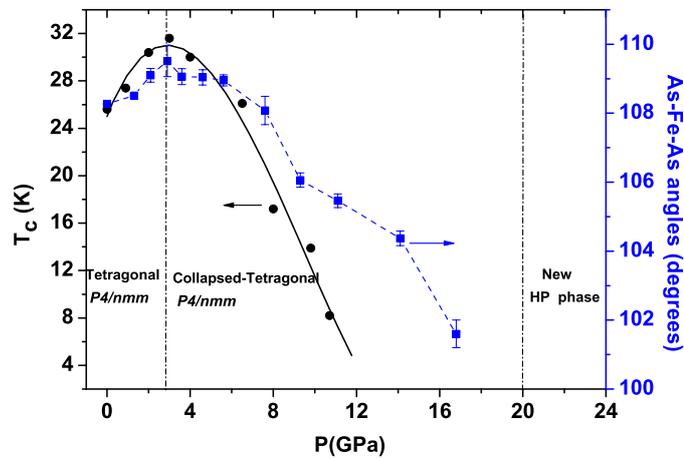

TOC

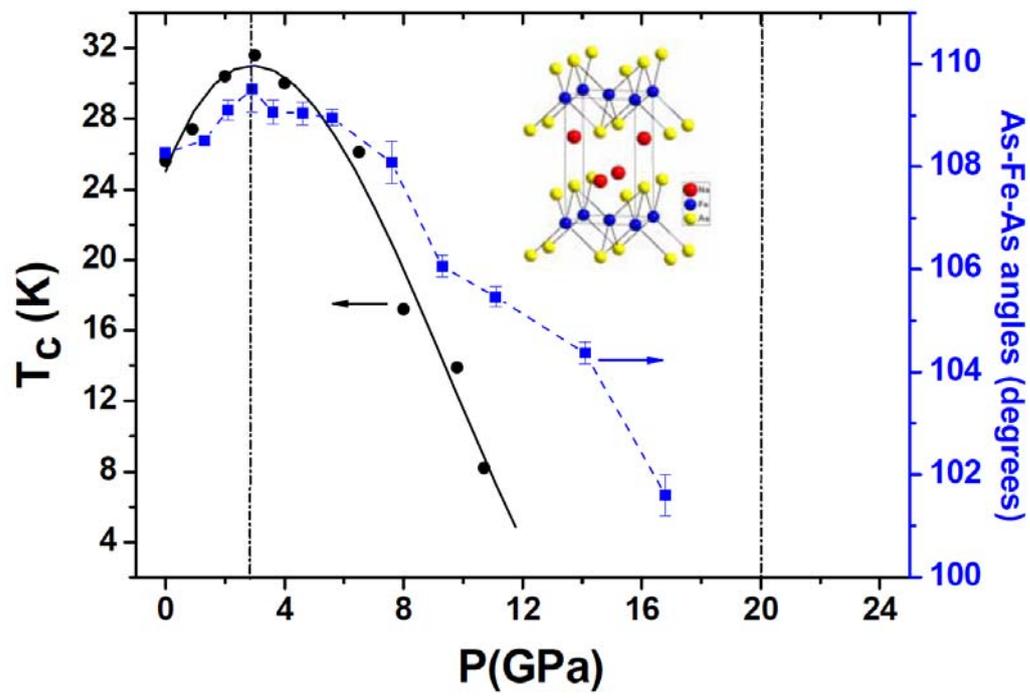